\documentstyle[12pt]{article}

\def\la{\langle}
\def\ra{\rangle}

\def\lb{\lbrack}
\def\rb{\rbrack}

 \def\Slash#1{
  \begin{picture}(5,6)(0,0)
  \put(-.7,-1.2){\line(5,6)6}
  \end{picture}
  \kern-.8em#1}
 \def\slash#1{
  \begin{picture}(5,6)(0,0)
  \put(-1.5,-1.7){\line(5,6)5}
  \end{picture}
  \kern-.8em#1}

\def\Sn{\Slash \nabla}
\def\sd{\Slash \partial}

\def\Tr{\mbox{Tr}}
\def\tr{\mbox{tr}}
\def\e{\epsilon}
\def\g5{\gamma_5}
\def\hg5{\hat{\gamma}_5}

\def\index{\mbox{index}\,}
\def\O{{\cal O}}
\def\F{{\cal F}}
\def\C{{\cal C}}
\def\H{{\cal H}}

\def\Qlatmr1{Q_{lat}^{(m=r=1)}}

\def\be{\begin{eqnarray}}
\def\ee{\end{eqnarray}}

\parskip=\medskipamount

\begin{document}

\vspace{8mm}

\begin{center}

{\Large \bf On the continuum limit of fermionic topological charge in
lattice gauge theory}
\\

\vspace{12mm}

{\large David H. Adams}

\vspace{4mm}

Math. dept. and Centre for the Subatomic Structure of Matter, \\
University of Adelaide, S.A. 5005, Australia. \\

\vspace{1ex}

email: dadams@maths.adelaide.edu.au

\end{center}

\begin{abstract}

It is proved that
the fermionic topological charge of $SU(N)$ lattice gauge fields on the
4-torus, given in terms of a spectral flow of the Hermitian Wilson--Dirac 
operator, or equivalently, as the index of the Overlap Dirac operator, 
reduces to the continuum topological charge in the classical continuum limit
when the parameter $m_0$ is in the physical region $0<m_0<2$.

\end{abstract}

Let $T^4$ denote the Euclidean 4-torus with fixed edge length $L$ and
fundamental domain $[0,L]^4\subset{\bf R}^4$. A gauge potential on an
SU(N) bundle over $T^4$ can be viewed as an su(N)-valued gauge field 
$A_{\mu}(x)$ on ${\bf R}^4$ satisfying
\be
A_{\mu}(x+Le_{\nu})=\Omega(x,\nu)A_{\mu}(x)\Omega(x,\nu)^{-1}
+\Omega(x,\nu)\partial_{\mu}\Omega(x,\nu)^{-1}
\label{1}
\ee
where $e_{\nu}$ is the unit vector in the positive $\nu$-direction and
$\Omega(x,\nu)\,$, $\nu=1,2,3,4\,$, are the SU(N)-valued monodromy
fields which specify the principal SU(N) bundle over $T^4$.\footnote{
These also satisfy a cocycle condition which ensures that 
$A_{\mu}(x+Le_{\nu}+Le_{\rho})$ is unambiguous and that eq. (\ref{8})
below is consistent. It is always possible to
make a gauge transformation so that $\Omega(x,\nu)=1$ for $\nu=1,2,3$
and $\Omega(x,4)$ is periodic in $x_1,x_2,x_3$. Then for fixed $x_4$
$\Omega(x,4)$ determines a map $T^3\to\mbox{SU(N)}$. The degree of this map
(which is independent of $x_4$ since $\Omega(x,4)$ depends smoothly on $x_4$)
equals the Pontryargin number of the SU(N) bundle over $T^4$.}
The Pontryargin number of the bundle is encoded in the gauge field as
its topological charge:
\be
Q=\frac{-1}{8\pi^2}\int_{T^4}\tr(F\wedge{}F)=
\frac{-1}{32\pi^2}\int{}d^4x\,\epsilon_{\mu\nu\rho\sigma}
\tr(F_{\mu\nu}(x)F_{\rho\sigma}(x))
\label{2}
\ee
The sections $\psi(x)$ in the standard spinor bundle over $T^4$ twisted
by the SU(N) bundle can be viewed as spinor fields on ${\bf R}^4$
satisfying
\be
\psi(x+Le_{\nu})=\Omega(x,\nu)\psi(x)
\label{3}
\ee
The Dirac operator $\sd^A=\gamma^{\mu}(\partial_{\mu}+A_{\mu})$ acts on 
these, and the Index Theorem \cite{AS} gives
\be
Q=\index\sd^A
\label{4}
\ee
The $\index\sd^A$ is equal to the spectral flow of the hermitian operator
$-\g5(i\sd^A-m)$ as $m$ increases from any negative to any positive
value (note that eigenvalues can only cross the origin at $m=0$ since
$(\g5(i\sd-m))^2=\sd^2+m^2$). 

The spectral flow description of $Q$ motivates a fermionic definition
of topological charge $Q_{lat}$ in lattice gauge theory 
\cite{Itoh,ov1,ov2}, which has been extensively studied numerically
in its various guises; see, e.g., 
\cite{Itoh,Smit,Seiler(Top),ov2,Vranas,Edwards,Gattringer,Hernandez,Chiu}
The purpose of this paper is 
to analytically prove that $Q_{lat}$ reduces to $Q$ in the classical 
continuum limit. (This result was announced in \cite{Taiwan} although the 
argument we give here is simpler and more direct than the one sketched
there.)

Put a hyper-cubic lattice on ${\bf R}^4$ with sites $a{\bf Z}^4$.
We consider only the lattice spacings $a$ for which $L/a$ is a
whole number. Furthermore we restrict to lattice spacings with the property
$a_1{\bf Z}^4\subset{}a_2{\bf Z}^4$ for $a_2<a_1$. This implies that if
$x\in{\bf R}^4$ is a lattice site in the lattice with spacing $a$
then it is also a lattice site in all the other lattices with spacing
$a'<a$. In the following, in statements concerning $a\to0$ limits 
(in particular Proposition 2 below) the variable $x$ always denotes
such a point in ${\bf R}^4$; it is fixed in ${\bf R}^4$ and does not
change as we go from one lattice to another.

The lattice transcript of $A$,
\be
U_{\mu}(x)=T\exp\Bigl(\,\int_0^1\,aA_{\mu}(x+tae_{\mu})\,dt\Bigr)
\label{5}
\ee
($T$ = $t$-ordering) satisfies
\be
U_{\mu}(x+Le_{\nu})=\Omega(x,\nu)U_{\mu}(x)\Omega(x+ae_{\mu},\nu)^{-1}\,.
\label{6}
\ee
Given such a lattice,
let $\C$ denote the infinite-dimensional complex vectorspace of lattice
spinor fields $\psi(x)$ (i.e. functions on the lattice sites taking
values in ${\bf C}^4\otimes{\bf C}^N$) and define the inner product
\be
\la\psi_1,\psi_2\ra=a^4\sum_{x\in{}a{\bf Z}^4}\psi_1(x)^*\psi_2(x)
\label{7}
\ee
where a contraction of spinor and colour indices is implied.
Let $\H\subset\C$ denote the Hilbert space of spinor fields with
$||\psi||<\infty$ and let $\C_L\subset\C$ denote the finite-dimensional
subspace of spinor fields satisfying the lattice version of (\ref{3}):
\be
\psi(x+Le_{\nu})=\Omega(x,\nu)\psi(x)\qquad,\qquad\forall\,x\in{}a{\bf Z}^4
\label{8}
\ee
The fields $\psi\in\C_L$ are determined by their restriction to 
$\F_L:=$the set of lattice sites contained in $[0,L)^4\subset{\bf R}^4$.
We define an inner product in $\C_L$ by
\be
\la\psi_1,\psi_2\ra_L=a^4\sum_{x\in\F_L}\psi_1(x)^*\psi_2(x)
\label{8.5}
\ee
The covariant forward (backward) finite difference operators
$\frac{1}{a}\nabla_{\mu}^+\,$ ($\frac{1}{a}\nabla_{\mu}^-$) are 
defined on $\C$ by
\be
\nabla_{\mu}^+\psi(x)&=&U_{\mu}(x)\psi(x+ae_{\mu})-\psi(x) \label{9} \\
\nabla_{\mu}^-\psi(x)&=&\psi(x)-U_{\mu}(x-ae_{\mu})^{-1}\psi(x-ae_{\mu}) 
\label{10}
\ee
These are bounded ($||\nabla_{\mu}^{\pm}||\le2$) and therefore map $\H$
to $\H$. They also preserve (\ref{8}) and therefore map $\C_L$ to
$\C_L$. Note that 
\be
(\nabla_{\mu}^{\pm})^*=-\nabla_{\mu}^{\mp}
\label{11}
\ee
on $\H$ and $\C_L$. The lattice version of $i\sd^A$ is the Wilson-Dirac
operator:
\be
D_w=i\frac{1}{a}\Sn+\frac{r}{2}a(\frac{1}{a^2}\Delta)\qquad\quad{}r>0
\label{12}
\ee
where $\frac{1}{a}\Sn=\sum_{\mu}\gamma^{\mu}\frac{1}{2}(\nabla_{\mu}^+
+\nabla_{\mu}^-)$ is the naive lattice Dirac operator and
$\frac{1}{a^2}\Delta=\frac{1}{a^2}\sum_{\mu}(\nabla_{\mu}^-+\nabla_{\mu}^+)
=\frac{1}{a^2}\sum_{\mu}(\nabla_{\mu}^+)^*\nabla_{\mu}^+=\frac{1}{a^2}
\sum_{\mu}(\nabla_{\mu}^-)^*\nabla_{\mu}^-$ is the lattice Laplace 
operator. Note that $\Sn$ is hermitian\footnote{We are following the 
maths convention where the $\gamma^{\mu}$'s are anti-hermitian.
This explains the factor $i$ in $i\frac{1}{a}\Sn$ in (\ref{12}) which is
not usually present in the physics literature where the $\gamma^{\mu}$'s
are hermitian.}
due to (\ref{11}) and $\Delta$ is hermitian and positive. 
(The Wilson term, i.e. the second term in (\ref{12}), which formally vanishes
in the $a\to0$ limit, is included to avoid the fermion doubling problem:
a degeneracy of the nullspace of $\Sn$ which is a lattice artifact unrelated
to the continuum theory \cite{Wilson,NN}.)
The lattice version of $\g5(i\sd-m)$ is the hermitian operator
$\frac{1}{a}H_m\,$:
\be
\frac{1}{a}H_m&=&\g5(D_w-\frac{rm}{a}) \label{13} \\
H_m&=&\g5(i\Sn+r({\textstyle \frac{1}{2}}\Delta-m))
\label{14}
\ee
It can be shown that the spectrum of $H_m$ is symmetric and without zero
for all $m<0$. Hence the spectral flow of $-H_m$ as $m$ 
increases from any negative value to some positive value $m_0$ is equal
to half the spectral asymmetry of $-H_{m_0}$ \cite{ov1,ov2}.
This suggests the following fermionic definition of the topological charge
of the lattice gauge field $U_{\mu}(x)\,$: 
\be
Q_{lat}=Q_{m_0}:=-\frac{1}{2}\Tr\Bigl(\frac{H_{m_0}}{|H_{m_0}|}\Bigr)
\label{15}
\ee
where $H_{m_0}$ is acting on $\C_L$. The spectral flow of $H_m$ was first
studied numerically in \cite{Itoh}. The definition (\ref{15}) arose
in the overlap formulation of chiral gauge theory on the lattice
\cite{ov1,ov2}. $Q_{m_0}$ also arises as an index:
$Q_{m_0}=\index(D_{m_0}):=\Tr(\g5|_{\ker{}D_{m_0}})$ where
$D=\frac{1}{a}(1+\g5\frac{H}{|H|})$ is the Overlap Dirac operator
\cite{Neu(PLB)}.

Unlike in the continuum case, the
spectral flow of $-H_m$ depends on the final value $m_0>0$ of $m$.
Numerical studies have shown that for reasonably smooth lattice gauge fields,
e.g. when $U_{\mu}(x)$ is the lattice transcript of a smooth continuum 
gauge field and the lattice is reasonably fine, the eigenvalue crossings
of $-H_m$ are localised around $m=0,2,4,6,8$ \cite{Itoh,Edwards}. 
Furthermore, when the lattice gauge field is the lattice transcript of
a continuum field the spectral flow due to crossings close to
$m=0$ was found to reproduce the continuum topological charge $Q$. 
In this paper we complement the previous numerical studies with the 
following analytical result:

\vspace{1ex}

\noindent {\it Theorem.} In the above setting, where $U_{\mu}(x)$ is the
lattice transcript (\ref{5}) and $m_0\not\in\{0,2,4,6,8\}$, there exists
an $a_0>0$ (depending on $A_{\mu}(x)$ and $m_0$) such that
\be
Q_{m_0}=I(m_0)Q\qquad\quad\mbox{for all lattice spacings $a<a_0$}
\label{16}
\ee
where

\begin{tabular}{l|c|c|c|c|c|}
          & $0<m_0<2$ & $2<m_0<4$ & $4<m_0<6$ & $6<m_0<8$ & 
$m_0\not\in[0,8]$ \\ 
\cline{2-6}
$I(m_0)\;$= & $1$     &  $-3$   & $ 3$    &  $-1$   &  $ 0$ \\
\end{tabular}
\be
\label{17}
\ee

\vspace{1ex}

\noindent {\it Remarks.} (i) The dependence on $m_0$ in 
(\ref{16})--(\ref{17}) coincides with that found in the above-mentioned
numerical studies with smooth lattice gauge fields. \hfill\break
(ii) The definition (\ref{15}) of $Q_{m_0}$ is only meaningful when 
$H_{m_0}$ does not have zero-modes. In the present case this is guaranteed
when $m_0\not\in\{0,2,4,6,8\}$ and $a$ is sufficiently small. 
Indeed, it is known that when $||1-U(p)||<\e$ for all lattice plaquettes
$p$, where $U(p)$ is the product of the link variables $U_{\mu}(x)$
around $p$, then there is a lower bound $H_{m_0}^2>b$, depending only on $\e$
and $m_0$, such that for fixed $m_0\not\in\{0,2,4,6,8\}$ $b>0$ when $\e$
is sufficiently small. This bound was established in \cite{local}
(and improved in \cite{Neu(bound)}) for the case where $0<m_0<2$ and can
be generalised to arbitrary $m_0\not\in\{0,2,4,6,8\}$ \cite{DA(bound)}.
In the present case, where $U_{\mu}(x)$ is the lattice transcript (\ref{5}),
we have
\be
1-U(p_{x,\mu\nu})=a^2F_{\mu\nu}(x)+O(a^3)(x)
\label{17.5}
\ee
leading to 
\be
||1-U(p)||\,\sim\,O(a^2)
\label{17.8}
\ee
Hence the above-mentioned lower bound $H_{m_0}^2>b>0$ holds for all 
sufficiently small $a$. Here and in the following $O(a^p)(x)$ denotes
a function on the lattice sites $x\in\F_L$ such that
the operator norm of $O(a^p)(x)$, considered as a multiplication operator 
on $\C$, satisfies $||O(a^p)(x)||\le{}a^pK$ for all $x\in\F_L$
where $K$ is a constant independent 
of $a$ and $x$. (In (\ref{17.5}) $O(a^p)(x)$ takes values in the space
of linear maps on ${\bf C}^N\,$; 
sometimes $O(a^p)(x)$ will just be a ${\bf C}$-valued
function of $x$, in which case we have $|O(a^p)(x)|\le{}a^pK$.)
We discuss the derivation of (\ref{17.5})--(\ref{17.8}), and other bounds
used in the following, in an appendix. 
In general, to conclude (\ref{17.8}) from (\ref{17.5})
we need the $O(a^3)(x)$ term to satisfy $||O(a^3)(x)||\le{}a^3K$ for all
$x\in{}a{\bf Z}^4$. For general gauge field $A_{\mu}(x)$ on ${\bf R}^4$
this holds when $||A_{\mu}(x)||$ and 
$||\partial_{\mu}A_{\nu}(x)||$ are bounded on ${\bf R}^4$ (cf. the appendix).
In the present case the condition (\ref{1}) generally results in divergence 
of $A_{\mu}(x)$ at infinity (for topologically non-trivial field).
Nevertheless we still have (\ref{17.8}) in this case:
it is a consequence of (\ref{6}) 
(note that $||U_{\mu}(x)||=1$ since $U_{\mu}(x)$ is unitary)
and the fact that the
$O(a^3)(x)$ term satisfies $||O(a^3)(x)||\le{}a^3K$
when $x$ is restricted to be in the fundamental domain $\F_L$.

The strategy for proving the theorem is to express 
$Q_{m_0}$ as the sum of a density:
\be
Q_{m_0}=a^4\sum_{x\in\F_L}q_L(x)
\label{18}
\ee
and show that
\be
q_L(x)=I(m_0)q^A(x)+O(a)(x)\qquad\quad(x\in\F_L)
\label{19}
\ee
where
\be
q^A(x)=\frac{-1}{32\pi^2}\e_{\mu\nu\rho\sigma}\tr{}F_{\mu\nu}(x)
F_{\rho\sigma}(x)
\label{20}
\ee
Then $\lim_{a\to0}Q_{m_0}=I(m_0)Q\,$, and since $Q_{m_0}$ is integer
it follows that $Q_{m_0}$ must coincide with $I(m_0)Q$  for small non-zero
$a$ as stated in the theorem.

To specify the density $q_L(x)$ in (\ref{18}) we introduce the following
definitions. We decompose $\C=\C^{sc}\otimes({\bf C}^4\otimes{\bf C}^N)\,$,
$\H=\H^{sc}\otimes({\bf C}^4\otimes{\bf C}^N)$ where $\C^{sc}\,$, $\H^{sc}$
denote the corresponding spaces of scalar lattice fields.
$\H^{sc}$ has the orthonormal basis 
$\{\frac{\delta_x}{a^2}\}_{x\in{}a{\bf Z}^4}$ where 
$\delta_x(y)=\delta_{xy}$. For linear operator $\O_{\H}$ on $\H$ we define
$\O_{\H}(x,y)=\frac{1}{a^4}\la\frac{\delta_x}{a^2},\O_{\H}
(\frac{\delta_y}{a^2})\ra\,$; this is a linear operator on 
${\bf C}^4\otimes{\bf C}^N$ satisfying
\be
\O_{\H}\psi(x)=a^4\sum_{y\in{}a{\bf Z}^4}\O_{\H}(x,y)\psi(y)
\qquad\quad\forall\,\psi\in\H
\label{21}
\ee
There is also an obvious decomposition $\C_L=\C_L^{sc}\otimes
({\bf C}^4\otimes{\bf C}^N)$ with $\C_L^{sc}$ having the basis
$\{\phi_x\}_{x\in\F_L}$ where $\phi_x(y)=\frac{1}{a^2}\delta_{xy}$
for $y\in\F_L$ and is extended to 
$a{\bf Z}^4$ in accordance with (\ref{8}):
\be
\phi_x(y+Ln)=\frac{1}{a^2}\Omega^{(n)}(x)\delta_{xy}\ \ ,\quad
\Omega^{(n)}(x)=\prod_{\nu}\Omega(x,\nu)^{n_{\nu}}\ ,\quad{}n\in{\bf Z}^4
\label{20.5}
\ee
For linear operator $\O_L$ on $\C_L$ we define 
$\O_L(x,y)=\frac{1}{a^4}\la\phi_x,\O_L\phi_y\ra_L$ for $x,y\in\F_L\,$; 
this is a linear operator on ${\bf C}^4\otimes{\bf C}^N$ satisfying
\be
\O_L\psi(x)=a^4\sum_{y\in\F_L}\O_L(x,y)\psi(y)
\qquad\quad\forall\,\psi\in\C_L\ ,\ x\in\F_L
\label{22}
\ee
The Cauchy--Schwarz inequality gives $||\O_{\H}(x,y)||\le\frac{1}{a^4}
||\O_{\H}||$ and $||\O_L(x,y)||\le\frac{1}{a^4}||\O_L||_L$.

The definition (\ref{15}) of $Q_{m_0}$ can now be rewritten as (\ref{18})
with
\be
q_L(x)=-\frac{1}{2}\tr\Bigl(\frac{H}{\sqrt{H^2}}\Bigr)_L(x,x)
\label{23}
\ee
where $H=H_{m_0}$ and the trace is over spinor and colour indices
(i.e. over ${\bf C}^4\otimes{\bf C}^N$). The strategy for deriving 
(\ref{19})--(\ref{20}) is now to relate $q_L(x)$ to $q_{\H}(x)$, defined
by replacing $\Bigl(\frac{H}{\sqrt{H^2}}\Bigr)_L$ by 
$\Bigl(\frac{H}{\sqrt{H^2}}\Bigr)_{\H}$ in (\ref{23}). (The latter is defined 
via the spectral theory for bounded operators on Hilbert space.)
This approach was suggested to me by Martin L\"uscher \cite{L(private)}.
The point is that (\ref{19})--(\ref{20}) are relatively easy to derive
for $q_{\H}(x)$; in fact this has essentially already been done in
previous works \cite{KY,DA,Fuji,Suz}. One potentially problematic aspect
with regards to these previous calculations is that in the present case
$A_{\mu}(x)$ can diverge for $|x|\to\infty$. However, we will get around this
by exploiting the locality property of $\Bigl(\frac{H}{\sqrt{H^2}}\Bigr)_{\H}$ 
\cite{local}, which will allow to replace 
$A_{\mu}(x)$ by a gauge field which vanishes outside a bounded region
of ${\bf R}^4$.

The relation between $q_L(x)$ and $q_{\H}(x)$ is as follows:

\vspace{1ex}

\noindent {\it Proposition 1.} 
\be
\Bigl(\frac{H}{\sqrt{H^2}}\Bigr)_L(x,y)
=\sum_{n\in{\bf Z}^4}\Bigl(\frac{H}{\sqrt{H^2}}\Bigr)_{\H}(x,y+Ln)
\,\Omega^{(n)}(y)\qquad(x,y\in\F_L)
\label{24}
\ee
where $\Omega^{(n)}(x)$ is defined in (\ref{20.5}). 
In particular, setting $y=x$ and substituting in (\ref{23}) we get
\be
q_L(x)=q_{\H}(x)-\frac{1}{2}\sum_{n\in{\bf Z}^4-\lbrace0\rbrace}
\tr\Bigl(\frac{H}{\sqrt{H^2}}\Bigr)_{\H}(x,x+Ln)\,\Omega^{(n)}(x)
\label{25}
\ee

\vspace{1ex}

\noindent {\it Proof.} We begin by deriving a relation between
$\O_L(x,y)$ and $\O_{\H}(x,y)$ for bounded operators $\O$ on $\C$ 
which leave $\C_L$ invariant. The proposition will then follow by exploiting
the fact that $\Bigl(\frac{H}{\sqrt{H^2}}\Bigr)_L$ and
$\Bigl(\frac{H}{\sqrt{H^2}}\Bigr)_{\H}$ can be simultaneously approximated 
by such operators. The approximation part is necessary since 
$\frac{H}{\sqrt{H^2}}$ is not a well-defined operator on
the whole of $\C\,$; the technicalities are related to the fact that
$\C_L\not\subset\H$, i.e. elements in $\C_L$ can have infinite norm.

Let $\O$ be a bounded operator on $\C$ which maps $\C_L$ to itself.
Then it follows from the above definitions and (\ref{20.5}) that,
for $x,y\in\F_L\,$,
\be
\O_L(x,y)&=&\frac{1}{a^4}\la\phi_x,\O\phi_y\ra_L
=\sum_{z\in\F_L}\phi_x(z)(\O\phi_y)(z) \nonumber \\
&=&\frac{1}{a^2}(\O\phi_y)(x)
=a^2\sum_{z\in{}a{\bf Z}^4}\O_{\H}(x,z)\phi_y(z) \nonumber \\
&=&\sum_{n\in{\bf Z}^4}\O_{\H}(x,y+Ln)\,\Omega^{(n)}(y)
\label{26}
\ee
We now exploit the fact \cite{local} that $\frac{1}{\sqrt{H^2}}$
has a power series expansion $\kappa\sum_{k=0}^{\infty}t^kP_k(H^2)$
norm-convergent to $\Bigl(\frac{1}{\sqrt{H^2}}\Bigr)_L$ and
$\Bigl(\frac{1}{\sqrt{H^2}}\Bigr)_{\H}$ on $\C_L$ and $\H$ respectively.
$P_k(\cdot)$ is a Legendre polynomial of order $k\,$; 
$||P_k(H^2)||\le1\,$; $t=e^{-\theta}\,$; the constants $\kappa,\theta>0$
depend only on the (strictly positive) lower and upper bounds on $H^2$
\cite{local}. (We are assuming that $a$ is sufficiently small so that
$H^2$ has a lower bound $b>0$ cf. remark (ii) above). Set
\be
P^{(N)}:=H\Bigl(\,\kappa\sum_{k=0}^Nt^kP_k(H^2)\Bigr)
\nonumber
\ee
For arbitrary finite $N$ this is a bounded operator on $\C$ which maps
$\C_L$ to itself. In light of (\ref{26}), to prove the proposition it 
suffices to show that 
$\Bigl(\frac{H}{\sqrt{H^2}}\Bigr)_L(x,y)-P_L^{(N)}(x,y)$ and
$\sum_{n\in{\bf Z}^4}\Big\lb\Bigl(\frac{H}{\sqrt{H^2}}\Bigr)_{\H}(x,y+Ln)
-P_{\H}^{(N)}(x,y+Ln)\Big\rb\Omega^{(n)}(y)$ both vanish in the $N\to\infty$
limit. The former is obvious. To show the latter it suffices to show that
$\sum_{n\in{\bf Z}^4}\sum_{k=N+1}^{\infty}t^k
||P_k(x,y+Ln)||$ vanishes in the $N\to0$ limit.
(We have set $P_k(x,z)=\lb{}P_k(H^2)\rb(x,z)$.)
For simplicity we show this for $y=x$ (the relevant case for (\ref{25}));
the argument in the general case is a straightforward generalisation.
Since $P_k(H^2)$ is of order $k$ in $H^2$, and $H$ couples only nearest
neighbour sites, we have $P_k(x,x+Ln)=0$ when $\frac{L}{a}\sum_{\mu}
|n_{\mu}|>2k$. Since $||P_k(x,z)||\le\frac{1}{a^4}||P_k(H^2)||\le{}a^4$
it follows that
\be
&&\sum_{n\in{\bf Z}^4}\sum_{k=N+1}^{\infty}t^k||P_k(x,x+Ln)|| \nonumber \\
&&\qquad\le\ \frac{1}{a^4}\,t^N
\Bigl(\,\sum_{n\in{\bf Z}^4\;,\;\frac{L}{2a}\sum_{\mu}
|n_{\mu}|\le{}N}\,\sum_{k=1}^{\infty}t^k\Bigr)
+\frac{1}{a^4}\Bigl(\,
\sum_{n\in{\bf Z}^4\;,\;\frac{1}{2a}\sum_{\mu}|n_{\mu}|>N}
t^{\Bigl(\frac{L}{2a}\sum_{\mu}|n_{\mu}|\bigr)}
\sum_{k=1}^{\infty}t^k\Bigr) \nonumber \\
&&\label{27}
\ee
The first sum over $n$ vanishes as $N^4t^N$ for $N\to\infty$, while the
second clearly vanishes for $N\to\infty$ since it is convergent for finite
$N$. This completes the proof of the proposition.

We now derive a small $a$ bound on the second term in (\ref{25}).
The facts that $P_k(x,x+Ln)=0$ for $\frac{L}{a}\sum_{\mu}|n_{\mu}|>2k$
and $||P_k(H^2)||\le1$ imply the following locality property of 
$(\frac{1}{\sqrt{H^2}})_{\H}$ \cite{local}:
\be
\Big|\Big|\Bigl(\frac{1}{\sqrt{H^2}}\Bigr)_{\H}(x,x+Ln)\Big|\Big|
&\le&||\kappa\sum_{k\ge\frac{L}{2a}\sum_{\mu}|n_{\mu}|}t^kP_k(x,y)|| 
\nonumber \\
&\le&\kappa{}\,t^{\Bigl(\frac{L}{2a}
\sum_{\mu}|n_{\mu}|\Bigr)}\sum_{k=0}^{\infty}t^k\frac{1}{a^4}
=\tilde{\kappa}
\frac{1}{a^4}\exp\Bigl(-\theta\frac{L}{2a}\sum_{\mu}|n_{\mu}|\Bigr)
\label{28}
\ee
where $\tilde{\kappa}:=\kappa/(1-e^{-\theta})$. 
For sufficiently small $a$ this gives
\be
\Big|\Big|\sum_{n\in{\bf Z}^4-\lbrace0\rbrace}
\Bigl(\frac{1}{\sqrt{H^2}}\Bigr)_{\H}(x,x+Ln)\Big|\Big|
&\le&\sum_{n\in{\bf Z}^4-\lbrace0\rbrace}\frac{\tilde{\kappa}}{a^4}\prod_{\mu}
\exp\Bigl(-\theta\frac{L}{2a}|n_{\mu}|\Bigr) \nonumber \\
&\le&\frac{\tilde{\kappa}}{a^4}\prod_{\mu}\left\lb2\int_{1/2}^{\infty}
\exp\Bigl(-\theta\frac{L}{2a}t_{\mu}\Bigr)\,dt_{\mu}\right\rb \nonumber \\
&=&\tilde{\kappa}
\Bigl(\frac{4}{\theta{}L}\Bigr)^4\exp\Bigl(-\frac{\theta{}L}{a}\Bigr)
\label{29}
\ee
The second inequality follows from the fact 
that $\int_{1/2}^{\infty}\exp(-\frac{\theta{}L}{2a}t)\,dt\,\ge\,
\exp(-\frac{\theta{}L}{2a}\Bigr)$ for sufficiently small $a$.
It now follows from (\ref{25}) that $q_L(x)=q_{\H}(x)+O(e^{-\rho/a})$ for
sufficiently small $a$. (This had already been noted by M. L\"uscher in
the abelian case in
\cite{L(abelian)} although the derivation was not provided there.)

To prove the theorem it now suffices to to show (\ref{19})--(\ref{20})
for $q_{\H}(x)$ instead of $q_L(x)$, i.e. to show
\be
q_{\H}(x)=I(m_0)q^A(x)+O(a)(x)\qquad\quad\mbox{for $x\in\F_L$}
\label{30}
\ee
To simplify the derivation we exploit the fact that $q_{\H}(x)$
is local in the gauge field \cite{local}. Because of this it suffices to
show (\ref{30}) in the case where $A_{\mu}(x)$ is replaced by another 
$SU(N)$ gauge field $\tilde{A}_{\mu}(x)$ on ${\bf R}^4$ with 
$\tilde{A}_{\mu}(x)=A_{\mu}(x)$ in a neighbourhood of $[0,L]^4$ and 
$\tilde{A}_{\mu}(x)=0$ outside a bounded region of ${\bf R}^4$.
Specifically, we can take $\tilde{A}_{\mu}(x)=\lambda(x)A_{\mu}(x)$
where $\lambda(x)$ is a smooth function on ${\bf R}^4$ equal to 1 on
$[-d,L+d]^4\,$ ($d>0$) and vanishing outside a bounded region.
To see this, let $H$ and $\tilde{H}$ denote the operators defined by
(\ref{14}) with lattice gauge fields $U$ and $\tilde{U}$ being the 
lattice transcripts (defined by (\ref{5})) of $A$ and $\tilde{A}$
respectively. Then, for small $a$, just as for $H^2$ we have 
$\tilde{H}^2>b>0$ and an expansion 
$\Bigl(\frac{1}{\sqrt{\tilde{H}^2}}\Bigr)_{\H}=\kappa\sum_{k=0}^{\infty}
t^k\tilde{P}_k$ where $\tilde{P}_k=P_k(\tilde{H}^2)$. Since $H$ and
$\tilde{H}$ only couple nearest neighbour sites, $P_k(H^2)$ and 
$P_k(\tilde{H}^2)$ can only couple a lattice site in $[0,L]^4$ to another
lattice site in $[0,L]^4$ via a site outside of $[-d,L+d]^4$ if
if $k\ge2(d/2a)$. Therefore $P_k(x,y)=\tilde{P}_k(x,y)$ for $x,y\in\F_L$
when $k<d/a\,$, and we find by an analogous argument to the one leading to
(\ref{28}) that, for $x,y\in\F_L\,$,
\be
\Big|\Big|\,\Bigl(\frac{1}{\sqrt{H^2}}\Bigr)_{\H}(x,y)
-\Bigl(\frac{1}{\sqrt{\tilde{H}^2}}\Bigr)_{\H}(x,y)\,\Big|\Big|
&\le&\kappa\sum_{k\ge{}d/a}^{\infty}t^k||P_k(x,y)-\tilde{P}_k(x,y)||
\nonumber \\
&\le&\frac{2\tilde{\kappa}}{a^4}\,e^{-\theta{}d/a}
\label{31}
\ee
This together with the ultra-locality of $H$ and $\tilde{H}$ implies
\be
q_{\H}(x)=\tilde{q}_{\H}(x)+O({\textstyle \frac{1}{a^4}}e^{-\rho/a})(x)
\qquad\quad\mbox{for $x\in\F_L$}
\label{32}
\ee
In light of this, the theorem now follows from (a special case of) the
following:

\vspace{1ex}

\noindent {\it Proposition 2.} Let $A_{\mu}(x)$ be a general smooth
$SU(N)$ gauge field on ${\bf R}^4$ with the property that
$||A_{\mu}(x)||\,$, $||\partial_{\nu}A_{\mu}(x)||\,$, and 
$||\partial_{\sigma}\partial_{\nu}A_{\mu}(x)||$ are all bounded.
Let $H=H_{m_0}$ be defined as in (\ref{14}) with the lattice gauge field
being the lattice transcript (\ref{5}) of $A_{\mu}(x)$. 
Then $q_{\H}(x)=-\frac{1}{2}\tr\Bigl(\frac{H}{\sqrt{H^2}}\Bigr)_{\H}(x,x)$
satisfies $q_{\H}(x)=I(m_0)q^A(x)+O(a)(x)$ for all $x\in{}a{\bf Z}^4$,
where $||O(a)(x)||\le{}aK$ for some constant $K$ independent of $x$ and
small $a$.

\vspace{1ex}

Clearly the gauge field $\tilde{A}_{\mu}(x)$ introduced above satisfies 
the conditions of the proposition (since it vanishes outside a bounded
region). Combining the proposition with (\ref{32}) then gives (\ref{30}),
proving the theorem.

To prove proposition 2 we use an integral representation to expand
$\frac{1}{\sqrt{H^2}}$ as a power series following 
\cite{DA,Taiwan}.\footnote{The suggestion to use an integral 
representation was made to me by M. L\"uscher.} (This gives a more
explicit power series expansion than the expansion in Legendre polynomials
\cite{local} discussed above.) 
Henceforth all operators are assumes to be acting on $\H$
and we drop the subscript ``$\H$'' in the notation.
Also, from now on $O(a^p)(x)$ denotes a term with $||O(a^p)(x)||\le{}a^pK$
for all $x\in{}a{\bf Z}^4$ (not just for $x\in\F_L$).
We first decompose
\be
H^2=L-V
\label{33}
\ee
where
\be
L&=&-\nabla^2+r^2({\textstyle \frac{1}{2}}\Delta-m_0)^2 \label{34} \\
V&=&=\;ir\frac{1}{2}\gamma_{\mu}V_{\mu}-\frac{1}{4}
[\gamma_{\mu},\gamma_{\nu}]V_{\mu\nu} \label{35}
\ee
with
\be
V_{\mu}&=&\frac{1}{2}[(\nabla_{\mu}^++\nabla_{\mu}^-)\,,
\sum_{\nu}(\nabla_{\nu}^--\nabla_{\nu}^+)]
\label{33a} \\
V_{\mu\nu}&=&\frac{1}{4}\lb(\nabla_{\mu}^++\nabla_{\mu}^-)\,,
(\nabla_{\nu}^++\nabla_{\nu}^-)\rb
\label{33b} 
\ee
As pointed out in \cite{local}, the norms of the commutators of the
$\nabla_{\mu}^{\pm}$'s are bounded by $max_p||1-U(p)||$. The bound
(\ref{17.8}) on $||1-U(p)||$ is valid when the conditions of proposition 2
are satisfied (cf. the appendix), hence
\be
||V||\;\sim\;O(a^2)
\label{36}
\ee
It follows that for small $a$ we have $||V||<b/2$ where $b$ is the lower
bound on $H^2$ mentioned earlier in remark (ii). This in turn implies
the lower bound $L>b/2>0$ for the positive operator $L$ in (\ref{34}).
Thus for sufficiently small $a$ the operator $L$ is invertible, 
$||L^{-1}||\cdot||V||<1$, and we can make the expansion
\be
\frac{H}{\sqrt{H^2}}
&=&H\int_{-\infty}^{\infty}\frac{d\sigma}{\pi}\,
\frac{1}{H^2+\sigma^2} \nonumber \\
&=&H\int_{-\infty}^{\infty}\frac{d\sigma}{\pi}\,
\Bigl(\,\frac{1}{1-(L+\sigma^2)^{-1}V}\Bigr) 
\Bigl(\,\frac{1}{L+\sigma^2}\Bigr)
\nonumber \\
&=&\int_{-\infty}^{\infty}\frac{d\sigma}{\pi}\,
\sum_{k=0}^{\infty}H(G_{\sigma}V)^kG_{\sigma}\,.
\label{37}
\ee
where $G_{\sigma}:=(L+\sigma^2)^{-1}$. Note that the $\gamma$-matrices
in (\ref{33}) are all contained in $V$. Since the trace of $\g5$
times a product of less than 4 $\gamma$-matrices vanishes, the 
$k=0$ and $k=1$ terms in (\ref{37}) give vanishing contribution to
$q_{\H}(x)$. On the other hand, the terms with $k\ge3$ satisfy the following 
bound:
\be
&&\Big|\Big|\,\int_{-\infty}^{\infty}\frac{d\sigma}{\pi}\,
\sum_{k=3}^{\infty}\lb{}H(G_{\sigma}V)^kG_{\sigma}\rb(x,x)\,\Big|\Big|
\nonumber \\
&&\qquad\le\;\frac{1}{a^4}||H||
\int_{-\infty}^{\infty}\frac{d\sigma}{\pi}\,
\sum_{k=3}^{\infty}||G_{\sigma}||^{k+1}||V||^k \nonumber \\
&&\qquad\le\;a^2K^3||H||\left\lb
\int_{-\infty}^{\infty}\frac{d\sigma}{\pi}\,\frac{1}{(b/2+\sigma^2)^4}
\right\rb\,\sum_{k=0}^{\infty}\Bigl(\frac{2}{b}a^2K\Bigr)^k
\label{38}
\ee
where we have used (\ref{36}) and the bounds 
$G_{\sigma}<(b/2+\sigma^2)^{-1}\le2/b$. This is $O(a^2)$ since the integral
and sum are finite and remain so in the $a\to0$ limit. Hence only the $k=2$
term in (\ref{37}) contributes in the $a\to0$ limit:
\be
q_{\H}(x)=q_{\H}^{(2)}(x)+O(a^2)(x)
\label{39}
\ee
where
\be
q_{\H}^{(2)}(x)=-\frac{1}{2}\int_{-\infty}^{\infty}\frac{d\sigma}{\pi}\,
\tr\lb{}HG_{\sigma}VG_{\sigma}VG_{\sigma}\rb(x,x)\,.
\label{40}
\ee

For lattice operators $\O$ which are polynomials in $\nabla_{\mu}^{\pm}$
we denote by $\O^{(0)}$ the operator obtained by setting $U=1$ in
(\ref{9})--(\ref{10}). Standard arguments give (cf. the appendix)
$||H-H^{(0)}||\sim{}O(a)$ and $||L-L^{(0)}||\sim{}O(a)$. The latter implies
$||G_{\sigma}-G_{\sigma}^{(0)}||\sim{}O(a)\,$; this follows from
$G_{\sigma}-G_{\sigma}^{(0)}=G_{\sigma}^{(0)}(L^{(0)}-L)G_{\sigma}$
since $G_{\sigma}$ and $G_{\sigma}^{(0)}$ are bounded from above by $2/b$
when $a$ is sufficiently small.
This allows us to replace $H$ and $G_{\sigma}$ by $H^{(0)}$ and 
$G_{\sigma}^{(0)}$ in (\ref{40}) at the expense of an $O(a)(x)$ term.
Furthermore we have $||[L^{(0)},V]||\sim{}O(a^3)$ (cf. the appendix).
This leads to $||[G_{\sigma}^{(0)},V]||\sim{}O(a^3)$ as follows:
The bound $||\nabla_{\mu}^{\pm}||\le2$ and triangle inequalities
lead to an $a$-independent upper bound $L<c$ which allows to expand
\be
G_{\sigma}=\Bigl(\frac{1}{c+\sigma^2}\Bigr)
\Bigl(\frac{1}{1-\frac{c-L}{c+\sigma}}\Bigr)
=\frac{1}{c+\sigma^2}\sum_{m=0}^{\infty}\Bigl(\frac{c-L}{c+\sigma^2}\Bigr)^m
\nonumber 
\ee
Now, since 
\be
||[(c-L^{(0)})^m,V]||\;\le\;m||[L^{(0)},V]||\cdot||c-L||^{m-1}
\;\le\;m(a^3K)(c-b/2)^{m-1}
\nonumber
\ee
we get
\be
||[G_{\sigma}^{(0)},V]||\;\le\;\frac{a^3K}{c^2}\sum_{m=0}^{\infty}(m+1)
\Bigl(\frac{c-b/2}{c}\Bigr)^m
\nonumber
\ee
and this is $\sim{}O(a^3)$ since the sum converges (since $0<b/2<c$).
Taking this into account in (\ref{40}), it follows from (\ref{39}) that
\be
q_{\H}(x)&=&-\frac{1}{2}\int_{-\infty}^{\infty}\frac{d\sigma}{\pi}\,
\tr\lb{}H^{(0)}V^2(G_{\sigma}^{(0)})^3\rb(x,x)+O(a)(x) \nonumber \\
&=&-\frac{1}{2}\tr\Big\lb{}H^{(0)}V^2
\int_{-\infty}^{\infty}\frac{d\sigma}{\pi}\,\frac{1}{(L^{(0)}+\sigma^2)^3}
\Big\rb(x,x)+O(a)(x) \nonumber \\
&=&\frac{-3}{16}\tr\lb{}H^{(0)}V^2(L^{(0)})^{-5/2}\rb(x,x)+O(a)(x)
\label{41}
\ee
Evaluating the trace over spinor indices we find 
(with $\nabla_{\mu}=\frac{1}{2}(\nabla_{\mu}^++\nabla_{\mu}^-)$)
\be
q_{\H}(x)&=&\frac{-3r}{16}\e_{\mu\nu\rho\sigma}\tr\Big\lb
(-\nabla_{\mu}^{(0)}(V_{\nu}V_{\rho\sigma}+V_{\nu\rho}V_{\sigma})
+({\textstyle \frac{1}{2}}\Delta^{(0)}-m_0)V_{\mu\nu}V_{\rho\sigma})
(L^{(0)})^{-5/2}\Big\rb(x,x) \nonumber \\
&&\quad+O(a)(x)
\label{42}
\ee
where $V_{\mu}$ and $V_{\mu\nu}$ are given by (\ref{33a})--(\ref{33b}).
Calculations give (cf. the appendix)
\be
\lb\nabla_{\mu}^{\pm},\nabla_{\nu}^{\pm}\rb\psi(x)
&=&(a^2F_{\mu\nu}(x)+O(a^3)(x))\psi(x\pm{}ae_{\mu}\pm{}ae_{\nu}) 
\label{45} \\
\lb\nabla_{\mu}^{\pm},\nabla_{\nu}^{\mp}\rb\psi(x)
&=&(a^2F_{\mu\nu}(x)+O(a^3)(x))\psi(x\pm{}ae_{\mu}\mp{}ae_{\nu}) \label{46}
\ee
These determine the relevant contributions of $V_{\mu}$ and $V_{\mu\nu}$
in (\ref{42}).

We now exploit the fact that there is a Fourier transformation on 
$\H^{sc}$ (=the space of scalar lattice fields with 
$||\phi||^2=\sum_{x\in{}a{\bf Z}^4}|\phi(x)|^2<\infty$); in particular
$\delta_x$ has the Fourier expansion
\be
\delta_x=\int_{-\pi}^{\pi}\frac{d^4k}{(2\pi)^4}\,e^{-ikx/a}\phi_k
\label{47}
\ee
where $\phi_k(y):=e^{iky/a}$. For a general operator $\O$ this leads to
\be
\O_{\H}(x,x)&=&\frac{1}{a^4}\la\frac{\delta_x}{a^2},\O\frac{\delta_x}{a^2}\ra
=\frac{1}{a^4}\int_{-\pi}^{\pi}\frac{d^4k}{(2\pi)^4}\,e^{-ikx/a}
\frac{1}{a^4}\la\delta_x,\O\phi_k\ra \nonumber \\
&=&\frac{1}{a^4}\int_{-\pi}^{\pi}\frac{d^4k}{(2\pi)^4}\,e^{-ikx/a}
(\O\phi_k)(x)
\label{48}
\ee
In the case where
\be
\O=\e_{\mu\nu\rho\sigma}
(-\nabla_{\mu}^{(0)}(V_{\nu}V_{\rho\sigma}+V_{\nu\rho}V_{\sigma})
+({\textstyle \frac{1}{2}}\Delta^{(0)}-m_0)V_{\mu\nu}V_{\rho\sigma})
(L^{(0)})^{-5/2}
\label{49}
\ee
a calculation using (\ref{33a})--(\ref{33b}) with (\ref{45})--(\ref{46})
gives
\be
(\O\phi_k)(x)=32\pi^2\,a^4\,\lambda(k;r,m_0)(q^A(x)+O(a)(x))\phi_k(x)
\label{50}
\ee
where
\be
\lambda(k;r,m_0)=
\frac{\prod_{\nu}\cos{}k_{\nu}\Bigl(-m_0+
\sum_{\mu}(1-\cos{}k_{\mu})-\sum_{\mu}\frac{\sin^2k_{\mu}}{\cos{}k_{\mu}}
\Bigr)}{\Big\lb\,\sum_{\mu}\sin^2k_{\mu}+
r^2(-m_0+\sum_{\mu}(1-\cos{}k_{\mu}))^2\Big\rb^{5/2}}
\label{51}
\ee
It follows from (\ref{42}) and (\ref{48}) that
\be
q_{\H}(x)=I(r,m_0)q^A(x)+O(a)(x)
\label{52}
\ee
where 
\be
I(r,m_0)=\frac{-3r}{8\pi^2}\int_{-\pi}^{\pi}d^4k\,\lambda(k;r,m_0)\,.
\label{53}
\ee
This integral was evaluated earlier in \cite{DA,Suz}. It was 
found to be independent of $r>0$ and a locally constant function of $m_0$
with values given by (\ref{17}). This completes the proof of proposition 2.

\noindent {\it Remark.} It is straightforward to generalise the results
of this paper to $SU(N)$ gauge fields on the $2n$-torus for 
arbitrary $n\ge2$ and to $U(1)$ gauge fields on the 2-torus.

Finally, following the suggestion of a referee, we emphasize that a key point
in this work is that it is the topological charge (i.e. the integrated Chern 
character) rather than the topological density that is shown to have the 
correct continuum limit. In this respect the treatment differs from all
earlier treatments which are essentially limited to small (hence topologically 
trivial) fields.

{\it Acknowledgements.} This work has benefited greatly from the input
of Martin L\"uscher, for which I thank him. I also thank Ting-Wai Chiu 
and Herbert Neuberger for discussions/correspondence.
The author is supported by an ARC postdoctoral fellowship.

\vspace{1ex}

\noindent {\bf Appendix}

\vspace{1ex}

In this appendix we recall, for completeness, certain standard facts 
concerning the lattice transcript of a smooth continuum gauge field
on ${\bf R}^4$ which lead to the bounds used in this paper.
The lattice transcript (\ref{5}) can be written as
\be
U_{\mu}(x)=\sum_{n=0}^{\infty}a^n\int_{0\le{}t_1\le{}\cdots\le{}t_n\le1}
dt_n\cdots{}dt_1\,A_{\mu}(x,t_n)\cdots{}A_{\mu}(x,t_1)
\label{A1}
\ee
where $A_{\mu}(x,t)=A_{\mu}(x+(1-t)ae_{\mu})$. When $A$ is bounded,
i.e. $||A_{\mu}(x)||\le{}K$ for all $x,\mu\,$, we have
\be
\Big|\Big|\sum_{n=p}^{\infty}a^n\int_{0\le{}t_1\le{}\cdots\le{}t_n\le1}
dt_n\cdots{}dt_1\,A_{\mu}(x,t_n)\cdots{}A_{\mu}(x,t_1)\,\Big|\Big|
&\le&\sum_{n=p}^{\infty}a^n\frac{1}{n!}K^n \nonumber \\
&\le&a^pK^pe^{aK}\,\sim\,O(a^p) \nonumber \\
&&\label{A2}
\ee
Therefore, to derive the $O(a^p)$ and $O(a^p)(x)$ bounds used in the text
it suffices to consider only a finite number of terms in the expansion
(\ref{A1}) (typically just the first few terms). An immediate consequence
of (\ref{A2}) with $p=1$ is the following: If $A$ is bounded then for 
any operator 
$P=P(\nabla_{\mu}^{\pm})$ which is a polynomial in the covariant finite
difference operators (\ref{9})--(\ref{10}) we have
\be
||P-P^{(0)}||\;\sim\;O(a)\,.
\nonumber
\ee
The bounds $||H-H^{(0)}||\sim{}O(a)$ and $||L-L^{(0)}||\sim{}O(a)$
are particular examples of this. If we furthermore assume that the first
order partial derivatives of $A$ are bounded, i.e. 
$||\partial_{\mu}A_{\nu}(x)||\le{}K$ for all $x,\mu,\nu\,$, we have
\be
||\lb\nabla_{\mu}^{\pm(0)},U_{\nu}\rb||\;\sim\;O(a)\,.
\label{A4}
\ee
To see this, note that
\be
\lb\nabla_{\mu}^{+(0)},U_{\nu}\rb\psi(x)
&=&(U_{\nu}(x+ae_{\mu})-U_{\nu}(x))\psi(x+ae_{\mu}) \nonumber \\
&=&\Bigl(a\int_{0\le{}t\le1}dt\,(A_{\nu}(x+ae_{\mu},t)
-A_{\nu}(x,t))+O(a^2)\Bigr)\psi(x+ae_{\mu}) \nonumber \\
&&\label{A5}
\ee
By the middle-value theorem,
\be
A_{\nu}(x+ae_{\mu},t)-A_{\nu}(x,t)=\partial_{\mu}A_{\nu}(x+sae_{\mu},t)
\nonumber
\ee
for some $s\in[0,1]$. Since $||\partial_{\mu}A_{\nu}||$ is bounded
(\ref{A4}) now follows from (\ref{A5}).
The bound (\ref{A4}) has the following easy generalisation:
Let $P=P(\nabla_{\mu}^{\pm})$ be a polynomial of degree $k$ in 
the $\nabla_{\mu}^{\pm}$'s; then if all the partial derivatives of
$A$ of order $\le{}k$ are bounded we have
\be
||\lb{}P^{(0)},U_{\nu}\rb||\;\sim\;O(a)
\label{A7}
\ee
Moreover, with the same boundedness assumptions on $A_{\mu}(x)$ and
$\partial_{\mu}A_{\nu}(x)$ straightforward calculations using the 
middle-value theorem give 
\be
1-U(p_{x,\mu,\nu})=a^2F_{\mu\nu}(x)+O(a^3)(x)
\label{A8}
\ee
Noting that \cite{local}
\be
\lb\nabla_{\mu}^+,\nabla_{\nu}^+\rb\psi(x)
=(1-U(p_{x,\mu\nu}))U_{\mu}(x)U_{\nu}(x+ae_{\mu})\psi(x+ae_{\mu}+ae_{\nu})
\label{A9}
\ee
and similar formulae for the other commutators, a straightforward refinement
of the arguments leading to (\ref{A7}) and (\ref{A8}) shows
\be
||\lb{}P^{(0)},\lb\nabla_{\mu}^{\pm},\nabla_{\nu}^{\pm}\rb\rb||\,\sim\,O(a^3)
\quad,\quad
||\lb{}P^{(0)},\lb\nabla_{\mu}^{\pm},\nabla_{\nu}^{\mp}\rb\rb||\,\sim\,O(a^3)
\label{A10}
\ee
The requirement for this is that $A$ and all its partial derivatives up to
order $r$ be bounded, where $r=min\{k,2\}$. Since $V$ is a linear combination
of commutators of the $\nabla_{\mu}^{\pm}$'s we have in particular
$||[L^{(0)},V]||\sim{}O(a^3)$ when
$A$ and its partial derivatives up to order 2 are bounded.
Finally, we remark that (\ref{45})--(\ref{46}) follow from combining
(\ref{A9}) and the corresponding formulae for the other commutators
with (\ref{A8}).

\end{document}